# Beyond the myth of the supernova-remnant origin of cosmic rays

Yousaf Butt

*High Energy Astrophysics Division, Harvard-Smithsonian Center for Astrophysics, 60 Garden Street, Cambridge, Massachusetts 02138, USA.*

**The origin of Galactic cosmic-ray ions has remained an enigma for almost a century. Although it has generally been thought that they are accelerated in the shock waves associated with powerful supernova explosions—for which there have been recent claims of evidence—the mystery is far from resolved. In fact, we may be on the wrong track altogether in looking for isolated regions of cosmic-ray acceleration.**

Somewhere out in space, cosmic rays are mysteriously being hurled to extreme energies. The fastest of them travel very close to the ultimate speed limit: the speed of light. These particles, mostly protons, but also other ions and electrons, permeate our Galaxy and rain down on earth continuously, night and day. Although cosmic rays were discovered almost a century ago, back in the balloon age, their origins remain unclear even now. Almost no effort has been spared in pursuing this long-standing mystery: satellites, rockets and balloons have been launched, and enormous detector arrays have been installed on the ground and even under mountains and seas. One remarkable detector array, called IceCube, is several times larger than the Eiffel Tower and is buried more than a kilometre beneath the clear Antarctic ice.

Cosmic rays are divided into two main classes according to their energy and probable acceleration sites: those below about $10^{18}$ eV in energy are called Galactic cosmic rays (GCRs); above that energy, they are referred to as extragalactic cosmic rays, although the exact demarcation energy remains somewhat vague and debatable, and there could be some overlap. By comparison, the most powerful man-made particle accelerator, the Large Hadron Collider near Geneva, will reach energies of ,$10^{13}$ eV, which is barely one-ten millionth of the most energetic cosmic ray recorded.

Here I exclusively discuss the lower-energy, Galactic, variety of cosmic rays. These particles are thought to be accelerated gradually, over centuries and even millennia, in the shock waves created by powerful supernova explosions within the Galaxy. As far back as 1953, Shklovskii speculated that "it is possible that ionized interstellar atoms are accelerated in the moving magnetic fields connected with an expanding [supernova remnant] nebula"[1]. However, we still have no proof of this scenario.

As GCR ions carry the bulk of the GCR energy, the main challenge rests in unambiguously identifying the origins of the ion, as opposed to the electron, component. Currently, the most direct way to find GCR acceleration sites is to look for telltale sources of $\gamma$-rays coincident with suspected celestial source sites. If an object is a GCR accelerator then it ought to have an overdensity of freshly accelerated cosmic rays in its vicinity. This 'cloud' of energetic cosmic rays can interact with the ambient matter and radiation to produce energetic $\gamma$-rays that can be detected on the Earth. The quandary is that both cosmic-ray ions and cosmic-ray electrons at the source site can produce the $\gamma$-rays we see, and it is extremely difficult to determine which type of particle was responsible for generating them.

There is fierce debate regarding the origin of the $\gamma$-rays seen in the direction of a few isolated supernova remnants (SNRs): whereas some advocate that ions are the source[2], others point out that electrons cannot be ruled out[3]. Though certainly fascinating, the outcome of this debate will not solve the puzzle of the origin of GCRs.

Even if we eventually find direct evidence that some isolated SNRs are accelerating ions, this will not automatically prove that such objects are the main sources of GCRs, in general.

In fact, there are already sufficiently serious flaws in the standard picture that cosmic-ray ions originate in isolated SNRs that it is now necessary to entertain alternative, more comprehensive and realistic, ideas. In my view, the real challenge is not just finding individual cosmic-ray acceleration sites—a given cosmic ray need not even have a unique discrete site as its origin—but, rather, determining the integrated cosmic-ray acceleration process. This probably involves the entire Galaxy and its extended halo, together with its ensemble of isolated, as well as overlapping, SNRs (called superbubbles), in what may be considered a single holistic acceleration 'site' [4].

# The standard model
## Shock acceleration in isolated SNRs.

The mechanism believed to be responsible for accelerating charged particles in an individual SNR is diffusive shock acceleration (DSA): such particles repeatedly scatter off magnetic turbulence on both sides of an SNR shock front, gaining speed as a result of the difference in the plasma velocities on either side of the shock. The greater the velocity difference, the greater the energy gained by the particle per shock crossing, and the larger the magnetic field (and turbulence), the higher the particle crossing frequency.

In the past few years, the HESS telescope array in Namibia and the CANGAROO array in Australia have discovered extended tera electronvolt (TeV, $10^{12}$ eV) $\gamma$-ray emission from at least four isolated shell-type SNRs: RX J1713.7-3946, Vela Junior (RX J0852.0-4622), RCW86 and SN 1006. Perhaps it is not coincidental that all four are relatively young, less than 2,000 yr old. In these SNRs, there is a close correlation between the morphology of non-thermal X-ray and TeV emissions that suggests a common origin for the fluxes, namely the electrons[3,5,6], but viable ion emission models also exist[2].

## Dynamical evidence for acceleration in SNRs.

Two main effects are expected if an SNR shell is accelerating cosmic-ray ions: the physical separation between the forward shock and the 'contact discontinuity' (or reverse shock) should be considerably reduced, and the temperature at the forward shock should be depressed. Indeed, in images of Tycho's SNR made by NASA's Chandra X-ray Observatory, the separation between the forward shock and the contact discontinuity is smaller than expected—unless a significant fraction of the explosion energy has gone into the acceleration of cosmic-ray ions[7,8]. These measurements, along with evidence that the forward-shock temperature in the young remnant 1E 0102.2-7219 is lower than that expected from the measured expansion velocity[9], are indirect 'dynamical' evidence for the acceleration of cosmic-rays ions by SNR shocks. However, although ions may be being accelerated and may take up a large fraction of the SNR mechanical energy in these remnants, such evidence does not necessarily mean that the ions are being accelerated to TeV energies and beyond. And even if they are, as may be the case for part of SNR RCW 86's shell[10], it does not follow that isolated SNRs are the main source of cosmic rays.

## Evidence for acceleration in old SNRs.

Intriguingly, direct spectral signatures of GCR acceleration may have recently been seen in a handful of older SNRs, such as IC 443[11,12], W 28[13] and perhaps also W 41[14] and the recently discovered G353.6-0.7[15]. Both the MAGIC (in the Canary Islands) and VERITAS (in Arizona) telescopes have observed TeV $\gamma$-ray emission in the direction of IC 443[11,12]. This may be the signature of locally accelerated ions interacting with an abutting molecular cloud[16,17] (Fig. 1). However, there is also an energetic pulsar wind nebula not too distant from the TeV source region in IC 443, and it could be the ions accelerated by the pulsar — rather than by the SNR shock wave — that are diffusing out and powering the TeV emission in the adjacent cloud[18]. The spectrum of IC 443 measured by the EGRET instrument on board NASA's Compton Gamma Ray Observatory also appears to show a 'pion-hump' feature at about 70 MeV (see fig. 4 of ref. 19) possibly indicating ion acceleration and interaction there, although the statistics of the 'detection' are marginal at best. It will be very interesting to see if the Italian Space Agency's AGILE satellite and NASA's Fermi Gamma-ray Space Telescope (previously called GLAST), which are both orbiting $\gamma$-ray observatories, confirm this feature with higher significance. Could it be that these senior citizens of the SNR population (,10[4]–10[5] yr old) have a larger role in accelerating GCR ions than do the youths?[20]

## Problems with the standard picture

The three strongest arguments supporting SNRs as possible cosmic ray ion sources remain indirect: the theoretical spectrum of particles undergoing DSA, that is, a power-law spectrum with index of 22, supposedly agrees with that deduced from observations; SNRs are among the few Galactic sources that can satisfy the large energetic requirement of powering cosmic rays; and GCR electrons are observed to be accelerated in SNRs.

The first argument is not particularly compelling. Any accelerator in which a fractional gain in energy by some particles is accompanied by a fractional loss in the number of the remainder yields such a power-law spectrum[21]. This is a common feature of particles escaping from most acceleration regions, regardless of the precise processes involved. In any case, a joint analysis of the propagation and composition of cosmic rays favours[22] a source spectrum with a power-law index of around -2.35, not -2.0. More precise, nonlinear, versions of DSA make this discrepancy even worse, as they predict a 'concave' source spectrum with an index of about -1.5 at high energies[23]. Notably, DSA also faces significant empirical difficulties in explaining interplanetary particle acceleration[24,25].

Neither, to address the second argument, are SNRs the only source of mechanical energy in space: there is sufficient power, for instance, in Galactic rotation, which could also be tapped (perhaps by means of magnetic reconnection[21] or spiral density shocks[26]) to power GCRs. Other, more novel, sources of power include jets from accreting neutron stars and black holes, $\gamma$-ray bursts and pulsar outflows. In any event, most of the SNR power injected into the Galaxy is not in the form of isolated SNRs, but rather in conglomerations of SNRs and massive stars (superbubbles), as outlined in the next section. If only individual SNRs were responsible for accelerating all cosmic rays, they would need to be extraordinarily efficient because they are so few. This distinction between SNRs and superbubbles would only be a minor issue if it were known that the putative process of cosmic-ray acceleration in superbubbles is the same as that thought to be at work in SNRs, but this remains far from certain.

As for the third point, the argument that SNRs are seen to be accelerating electrons does not automatically mean they are necessarily important sources of GCR ions. Because energetic (>100-GeV) electrons lose energy much more rapidly than ions, sources of the GCR electrons seen on Earth are expected to be located, predominantly, within just a few kiloparsecs, whereas the acceleration regions of GCR ions are not similarly confined. (The entire Galaxy is about 30 kpc across.) GCR ions and electrons may have altogether different origins and acceleration processes.

The elemental and isotopic make-up of GCRs gives us further clues as to where they may originate. Such composition analysis indicates that they are accelerated mainly from a pool of old[27] ($>10^5$-yr-old), well-mixed interstellar material that does not reflect the elemental anomalies of fresh SNR ejecta[28]. In fact, isotopic anomalies in the GCR composition, such as the enhanced $^{22}Ne/^{20}Ne$ ratio, indicate[29] that GCRs are preferentially accelerated out of the material found in superbubbles.

The distribution of SNRs as a function of galactocentric distance also appears to be inconsistent with diffuse $\gamma$-ray data: the predicted cosmic-ray gradient, if due to SNRs alone, is steeper than that deduced from low-energy (<10-GeV) γ-ray maps. However, there may be ways to circumvent this problem[22]; for example, if more hydrogen gas exists farther from the centre of the Galaxy than is now thought, this could help redress this seeming inconsistency by compensating for the lower density of SNRs there.

Finally, the fact that GCRs are seen as coming from all directions with virtually equal intensity (that is, a 'low anisotropy', to better than 1 part in 1,000 at ~$10^{14}$ eV per nucleon) poses an even more serious challenge to the standard GCR origin picture. Such high energy GCRs would be expected to escape from the Galaxy relatively quickly and, therefore, result in a greater anisotropy than has been observed, were they accelerated only by individual SNRs[30–32]. Isolated SNRs can be playing, at best, only a minor part in accelerating such cosmic-ray ions.

Thus both low-energy (<10-GeV) γ-ray data and high-energy (~100-TeV) cosmic-ray data argue against a supernova remnant origin of GCRs. Most of the power needed to accelerate GCRs may be supplied by SNRs, but the acceleration mechanism appears to be more complex and is probably distributed throughout the Galaxy and its extended halo[4,30–35].

Last year's intriguing report by the Milagro collaboration of a slight excess (a few parts in 10,000) of 10-TeV cosmic rays coming from two patches of the sky[36] does not detract from this conclusion. It remains to be seen whether the Milagro excess means that the solar system is being bathed in excess cosmic-ray nuclei from a nearby cosmic-ray accelerator[37], whether it is simply a local effect having to do with the sun's magnetic field structure[36] or whether it is something else entirely.

# Distributed acceleration
### Superbubbles.

Most supernovae are of the core-collapse variety, having massive progenitor stars. Such massive stars are born in clusters of up to several thousand members, burn fast and die young, in near-simultaneous supernova explosions. The resulting multiple SNRs meld into one another, forming enormous superbubbles in the interstellar medium (ISM). Because most of the power injected by SNRs into the ISM is injected through such superbubbles, rather than through the isolated SNRs[38], it is imperative that we understand the role of superbubbles in GCR acceleration.

Unfortunately, the radiative signature of SNRs in the rarefied and hot medium of a superbubble interior is expected to be minimal[39], so we may be ignorant of most of the SNRs in the Galaxy. (These hidden SNRs might also explain the 'missing-SNR' problem, namely that we ought to see many more SNRs than have been detected so far.) Although our vantage point within the Galaxy makes it hard for us to observe and analyse the large and diffuse Galactic superbubbles, indirect evidence of GCR acceleration in superbubbles in the nearby Large Magellanic Cloud galaxy has recently been presented[40].

The current generation of TeV $\gamma$-ray telescopes could possibly detect some of the hidden SNRs in the Galaxy. Indeed, some of the many 'dark' extended TeV sources detected by the HEGRA and HESS collaborations could be such SNRs. The fact that the Milagro water-Cherenkov-detector group has detected extended $\gamma$-ray emission above 10 TeV coincident with some of these dark TeV sources makes them especially intriguing and worthy of pursuit as plausible GCR acceleration hotspots[41].

Large-scale putative GCR accelerators, such as superbubbles, would imply a spatially variable GCR intensity through the Galaxy, and the diffuse $\gamma$-ray maps made from data collected by the EGRET instrument do show possible evidence of this. Contiguous large scale $\gamma$-ray features, uncorrelated with known Galactic molecular gas concentrations, have been detected and are significantly brighter than the average $\gamma$-ray background (ref. 42 and Supplementary Fig. 2 therein). Although these features have sometimes been interpreted in terms of a mysterious Galactic 'dark gas'[42] (not to be confused with dark matter or dark energy), an alternative explanation is that they may be simply a consequence of enhanced GCR source intensity there.

A striking example is the bright and extended excess-c-ray region[42] coincident with the Gum nebula[43]. The Gum nebula is known to be internally powered by OB associations and/or SNRs[44]. A natural explanation for the excess diffuse $\gamma$-ray emission seen in this direction is that the cosmic-ray intensity there is significantly higher than the local one. Similarly, enhanced extended $\gamma$-ray emission is also coincident with a newly discovered superbubble in the constellation Ophiuchus[45]. It is most likely that the mysterious dark gas has different explanations in different regions of the sky: a spatially varying GCR intensity, recently detected molecular material not included in earlier models and perhaps some cold HI gas.

## Galaxy-wide acceleration.

Superbubbles cannot be the entire solution to the origin of GCRs, however, as they run into some of the same problems plaguing isolated SNRs. Superbubbles also cannot account for the low large-scale anisotropy of cosmic rays, nor explain the shallow cosmic-ray gradient deduced from $\gamma$-ray data. Even if isolated SNRs and superbubbles are considered the main power source for GCR acceleration, it is probable that the process is actually distributed across the Galaxy and the extended halo[30,31,33]. Cosmic ray reacceleration may also be taking place at the Galactic-wind termination shock at a distance of a few hundred kiloparsecs from the centre of the Milky Way[34], as well as in 'slipping interaction regions' (50–100 kpc distant) in the Galactic wind[35]. The collective reacceleration of the cosmic-ray particles by this shock ensemble may also explain the observable cosmic-ray spectrum up to energies of ,$10^{17}$ eV, as well as the low anisotropy of high-energy GCRs[31].

## Future prospects

The problem of the origin of cosmic rays is not that we have not yet found a firm spectral signature of ion acceleration in even a single isolated SNR, but that there are other, more severe, problems with this oversimplified scenario to begin with. Even if such a signature were found, it would not be sufficient to prove that isolated SNRs are the main accelerators of GCRs.

Because the process of GCR acceleration could be distributed[31,33], we may not even be posing the correct questions in trying to identify only discrete GCR source sites. A given cosmic ray need not have originated from exactly one source: its 'origin' may be intrinsically unclear.

What is needed is a better integration of 'microscopic' source-specific discrete acceleration models, with macroscopic Galactic—and extended halo, plus termination shock—propagation 22 and reacceleration 31 models, to provide a comprehensive picture of how GCRs gain energy. An early prototype of such an integrated model is the study in ref. 33. Although more sophisticated numerical models have since been developed (for example the popular GALPROP code 22), important shortcomings remain. For instance, owing to the size of its numerical grid, GALPROP is currently unable to properly reproduce fine-scale spatial and temporal variations expected from localized sources of GCRs 46. Much information about the origin of cosmic rays remains to be uncovered by modelling the Galactic 'ecology' of GCR acceleration, reacceleration, transport and composition holistically and at high fidelity.

A great deal of theoretical and observational work remains ahead. It would be useful, for instance, to understand whether the putative process of particle acceleration in superbubbles (that is, multiple interacting shocks embedded in pre-existing strong turbulence) could be as—or, perhaps, even more—efficient than that thought to operate in isolated SNR shocks. Supernovae provide the main energy source for superbubbles, but the details of the respective acceleration mechanisms in superbubbles and SNRs may be quite different. If, for example, the explosion energy is dumped into magnetic turbulence in the interior of a superbubble, it is conceivable that the superbubble, as a whole, acts as an accelerator with the second-order Fermi process (that is, stochastic acceleration) dominating, and with the potential for a large increase in maximum cosmic-ray energy because of the increase in spatial scale. On the other hand, if the energy stays in individual SNR blast waves, the regular DSA mechanism will dominate, with acceleration preferentially occurring across the superbubble wall. The resolution to the mystery of the origin of GCRs will be incomplete until we have a better understanding of the role of superbubbles in GCR acceleration.

Observations of superbubbles will be equally important: for example, are any Galactic superbubbles $\gamma$-ray bright, indicating possible cosmic-ray acceleration there? Is there any evidence for nonthermal emission from them, as there is for several Large Magellanic Cloud superbubbles[40]? Are some of the extended hotspots attributed to the mysterious dark gas in diffuse $\gamma$-ray emission maps (ref. 42 and supplementary fig. 2 therein) really due to superbubbles? Are some of the dark TeV sources related to SNRs otherwise invisible because they are exploding within superbubbles[39]? Forthcoming maps of diffuse $\gamma$-ray emission recorded by the AGILE and Fermi observatories will go a long way towards answering some of these questions. Further in the future, neutrino observatories will also have an important role in understanding just how GCRs are accelerated.

Nevertheless, we should be prepared for the inevitable complications: for example, how do we find weak, diffuse GCR accelerators if the whole galaxy is aglow in $\gamma$-rays (or neutrinos), as it is? How important are discrete acceleration sites in comparison with distributed acceleration and reacceleration? How are we to discriminate between the $\gamma$-rays coming from extended acceleration regions and those arising from cosmic-ray propagation and interaction, in a Galaxy with a spatially variable cosmic-ray intensity? Is the usual assumption that the parameters of the cosmic-ray flux measured near Earth apply Galaxy-wide really correct? Unless we start asking the difficult questions, cosmic rays may hold fast to the secret of their origin for another century.

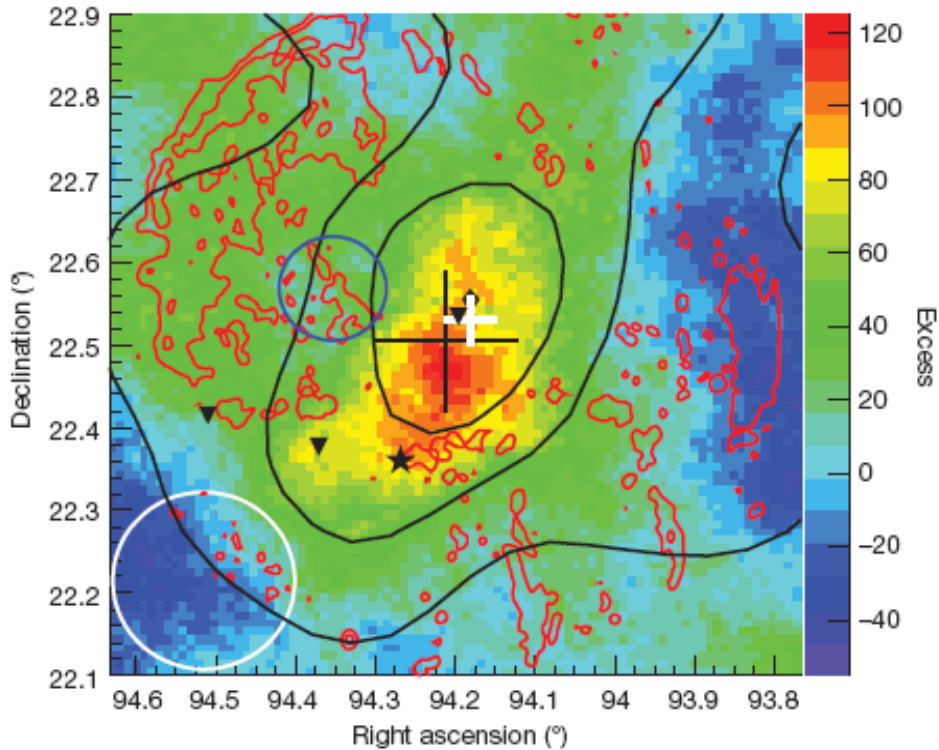

Figure 1 | Possible γ-ray emission from an SNR shock wave. An image, generated from VERITAS c-ray telescope data, showing the very high-energy γ-ray signal detected in the direction of the SNR IC 443 (colour scale)[12]. The black cross indicates the centroid position of the extended TeV γ-ray emission, and its uncertainty. Similarly, the white cross indicates the position and uncertainty of the MAGIC telescope's TeV source[11] MAGIC J0616 1225. The optical emission contours give an indication of the overall size of the SNR and are shown in red. The blue circle indicates the 95% error circle of the nearby lower-energy Fermi γ-ray source 0FGL J0617.4 1 2234. The intensity of the carbon monoxide emission, which is a proxy for the approximate amount of molecular cloud material in the vicinity of the SNR, is shown in black. The coincidence of TeV emission with the molecular material may be indicative of locally accelerated cosmic rays interacting with the ambient material. The locations of maser emission (possibly indicative of shock–cloud interactions) are shown by triangles, and the pulsar CXOU J061705.3 1222127 is indicated by a star shape. The white circle illustrates the point spread function of the VERITAS telescope. For further details, see ref. 12. (Figure produced by B. Humensky for VERITAS[12]; reproduced by permission of the American Astronomical Society.)

This study suggests that distributed Galaxy-wide acceleration could be an acceptable mechanism for Galactic cosmic-ray acceleration.

Acknowledgements: Part of this work was carried out while the author was a fellow at the National Academy of Sciences. The support of a NASA Long Term Space Astrophysics grant is gratefully acknowledged.

Author Information Correspondence should be addressed to Y.B. (ybutt@cfa.harvard.edu).